# Large-Angle, Single-Order, One and Two-Dimensional Atomic Interferometry for Creation of Nanostructures

M.S. Shahriar and T. Zelevinsky
*Research Laboratory of Electronics, Massachusetts Institute of Technology, Cambridge, MA 01239*

P.R. Hemmer
*Air Force Research Laboratory, Hanscom AFB, MA 01731*

**Abstract**

We have shown via explicit analysis as well as numerical simulation the design of two large angle interferometers employing two-photon pulses. The first one uses the technique of adiabatic following in a dark state to produce a large splitting angle atomic interferometer that is capable of producing one dimensional gratings with spacings as small as 2 nm. Unlike other large angle interferometers, this technique is not sensitive to errors in optical pulse area and decoherence from excited state decay. This may lead to a nearly two orders of magnitude improvement in the sensitivity of devices such as atomic gyroscopes, which are already as good as the best laser gyroscopes. The second interferometer uses the technique of Raman pulses to produce a *two-dimensional interferometer*, with independent choice of grating spacings in each direction, each being as small as 2 nm. This scheme may enable one to produce uniform arrays of quantum dots with dimensions of only a few nm on each side. In addition, it may be possible to generalize this process to produce arbitrary patterns with the same type of resolution.

In recent years, rapid progress has been made in the area of atom interferometry. Atomic beam gyroscopes have been demonstrated with sensitivities exceeding that of ring laser gyroscopes [1,2]. Such a rotation sensor may enable one to measure the general relativistic Lens-Thirring rotation in the near future. Atom interferometers have also been used to measure precisely the ratio of Planck's constant to atomic mass, acceleration due to gravity, as well as the gradient thereof [3,4]. In some of these experiments, such as the gyroscope, the sensitivity is proportional to the area enclosed by the different paths, which in turn is determined in practice by the degree of splitting at the input port. For the current schemes, the splitting corresponds typically to a few photon recoils. As such, there is a need for an interferometer based on a much larger splitting angle . Using a high coherence-length source, such as the Bose-condensed atom laser [5] this type of interferometer will also enable the creation of one and two dimensional structures with feature sizes of less than 10 nm. With chemical substitution techniques, these structures can be transferred to semiconductors or coinage metals, yielding uniform arrays of nano-quantum dots, for example.

Several schemes have been studied for realizing large angle beam splitters. The interaction of a two level atom with a standing wave light field can produce large splitting; however, the atoms are scatters into multiple orders [6] because of the sinusoidal nature of the phase grating. The magneto-optic beam-splitter [7-9] and variations thereof [10,11] produces a triangular phase grating, which represents an improvement over the pure standing wave. Because of the subwavelength extent of the triangular shapes, however, the number of higher orders is still significant. In another scheme involving bichromatic standing waves exciting a two level transition, the potentials remain triangular for all the dressed states, and extend over many wavelengths[12]. Both of these schemes suffer from the problem that typically the atoms are in the excited state for nearly half the time. As such, in order to minimize decoherence, the interaction time has to be small compared to the natural lifetime. This in turn limits the maximum, coherent splitting to less than 20 photon recoils due to constraints imposed by high frequency modulators.

In this article, we propose a new beam splitter which can achieve a splitting exceeding ±100 photon recoils, and can be recombined easily to yield a high sensitivity interferometer. For concreteness, we consider the $^{87}$Rb atom, released from an evaporatively cooled magnetic trap (or a Bose condensate) and falling under gravity. The relevant energy levels are shown in figure 1. The atoms are assumed to be in the state |F=1, $m_f$=1> at the onset. Here, the quantization direction, **z**, is assumed to be normal to the direction of gravity, denoted as **y**.

Using a pair of linearly polarized laser beams, co-propagating in the **x** direction and polarized in the **z** direction, we excite the Raman transition, coupling |F=1, $m_f$=1>≡|a> to |F=2, $m_f$=1>≡|c>, via two π transitions (dashed-line transitions in Figure 1). The beams are detuned strongly from the excited manifold of the $D_2$ line, but are two-photon resonant, so that the process can be thought of as a two-level transition between the two magnetic sublevels. The duration of these beams are controlled precisely, corresponding to a π/2 pulse, putting the atom into the state (|a>-|c>)/√2. Alternatively, a counter-intuitive pulse sequence can be used, starting with only $D_2$ light and gradually

increasing the intensity of the $D_1$ light until the desired superposition state is reached via adiabatic passage, after which the fields are switched off. In this case, a large detuning from the excited state is not required. Ignoring the small difference between the photon recoils corresponding to the two legs of the Raman transition, we note that the center-of-mass momentum remains unchanged during this process.

The component of the atom remaining in the state |a> can now be deflected, in the –**z** direction, by using a set of counter-intuitively sequenced Raman pulses that couple this superposition state to $|F=1, m_f=-1\rangle \equiv |b\rangle$, via the intermediate state of $|F'=1, m_f=0\rangle$ of the $D_1$ manifold (solid-line transitions in Figure 1) [13,14]. Here, the optical beams are circularly polarized, and propagate in the ±**z** direction. Consider the effect of the first pair of pulses seen by the atom. The timing of the pulses, each of duration 2T, is controlled to ensure that the $\sigma_+$ polarized pulse, propagating in the +**z** direction arrives first, at time t=0. Halfway through this pulse (i.e., at t=T), the second pulse with $\sigma_-$ polarization and propagating in the –**z** direction, arrives at the atom. From t=0 to t=3T, the atom evolves adiabatically, always staying in the dark state corresponding to the Raman transition: $|D\rangle \propto \{g_-|a,0\rangle - g_+|b,-2\hbar k\rangle\}$, where $g_\pm$ are the Rabi frequencies corresponding to the $\sigma_\pm$ pulses, and k is the wave number corresponding to the optical transitions. As such, at t=3T, the atom is in equal superposition of $|c,0\rangle$ and $|b,-2\hbar k\rangle$, i.e, the part in the internal state |c> has no momentum in the **z** direction, while the part in the internal state |b> has a momentum corresponding to two photon recoils in the –**z** direction. The action of the first pair of pulses is illustrated by the solid line transitions in Figure 2, where for clarity the σ+ and σ- transitions are drawn as if they have different energies and energy shifts due to kinetic energy are omitted. The requirement for this process to remain adiabatic is that the average Rabi frequency be much larger than the inverse of the pulse duration; this condition can be easily satisfied for pulses as short as several 10's of nanoseconds. Furthermore, note that the pulse area for this process does not have to be exact in order for the transfer to be exact.

This process is now repeated during the second pair of pulses, where the atom sees first a $\sigma_-$ pulse moving in the **z** direction, followed an interval T later by a $\sigma_+$ pulse moving in the –**z** direction. After this pulse sequence, the atom ends up in equal superposition of $|c,0\rangle$ and $|a,-4\hbar k\rangle$. More generally, after 2N pairs of pulses alternated in this form, the atom is in an equal superposition of $|c,0\rangle$ and $|a,-4N\hbar k\rangle$, as shown in Figure 2 (long dashed-line transitions). Figure 3 shows the results of a numerical simulation for 4N=60. Here, the curve shows the mean center of mass momentum of the |a> component, as a function of the total interaction time. Note that as the atoms start gaining more and more momentum, they will tend to get out of single-photon resonance (two-photon detuning would alternate depending on the propagation directions of the σ+ and σ- beams). In this simulation, detuning has been compensated exactly by adjusting the frequency of the pulses; the same can be realized easily in an experiment as well. As such, this process may be characterized as a step-wise frequency chirped single order deflection, and, unlike most other beam-splitters/deflectors, is not bounded by the so-called Raman-Nath limit. Here, we have kept N limited to a relatively small number primarily because of computational constraints; in order to keep track of the momentum spread, the size of the density matrix scales as $N^2$, and the propagator for the density matrix scales as $N^4$. From this result, it can be inferred that the process can continue coherently for larger N, yielding momentum transfers as high as a few hundred recoils before the momentum spread would start becoming significant.

For the rubidium example, let us consider a situation where 2N=50, so that at the end of the pulse sequence, the atom is in an equal superposition of $|c,0\rangle$ and $|a,-100\hbar k\rangle$. We assume 3T=150 nsec, the time required for each pair of pulses to interact with the atom, and set g, the rms rabi frequency to 100 MHz, so that the adiabaticity parameter of $(2\pi g T)^{-1}$ is about 0.03, which is much less than unity, as required. The total duration for the splitting sequence is thus about 7.5 μsec. The gravitational drop during this interval is negligible compared to the size (about 1 mm) of the initial atomic cloud. Because of the velocity difference between the two components of the atomic state, the cloud will now separate spatially while falling under gravity. For $^{87}$Rb atoms, the momentum difference between these two components corresponds to a transverse velocity difference of about 0.6 m/sec. As such, in 3.3 msec, for example, the atomic cloud will separate spatially, in the **z** direction, by a distance of 2 mm, while each component individually will spread much less, since the velocity spread in the initial cloud is less than a single recoil. The corresponding drop in the **y** direction, due to gravity, will be only 55 μm.

At this point, another sequence of pulses, with 2N=100, is applied to the pulses, again along the **z** direction, with the direction of the pulses reversed, so that at the end of the pulse sequence the atom will be in an equal superposition of $|c,0\rangle$ and $|a,+100\hbar k\rangle$. This process will take only about 15 μsec, so that the separation between the cloud will remain essentially unchanged. We now apply a set of linearly polarized laser beams copropagating in the **x** direction, only to the $|c,0\rangle$ component of the atoms. Because of the large spatial separation between the clouds, this is easily possible. As before, these beams will excite off-resonant π transitions on the $D_2$ manifold, which will correspond to a resonant two-photon transition coupling $|c,0\rangle$ to $|a,0\rangle$. The duration of these beams is chosen to correspond to a π pulse. Therefore, the state of the atom will now be an equal superposition of $|a,0\rangle$ and $|a, +100\hbar k\rangle$, separated spatially by 2 mm.

These two clouds will now come together at the rate of about 60 cm/sec, spatially superimposing each other in about 3.3 msec, while dropping under gravity by about another 160 μm. The interference between these two arms will yield matter wave fringes, with a peak-to-peak spacing of about 8 nm. Using techniques that have already been well established [15] this pattern can first be deposited on a substrate coated, for example, with self-assembled monolayers of octyltrichlorosilane. The damage induced on this layer can then be transferred chemically to an underlying layer of semiconductors as well as coinage metals. This approach is obviously the preferred one for lithographic applications. Figure 4 summarizes the various steps involved in this process.

For interferometric applications, large path separations are required and Ramsey techniques are desired to detect the interference. To generate a large average path-separation (or large area) interferometer, the drift time can be increased. For example, taking a 50 msec drift time between pulse sequences gives a maximum separation of

3 cm in the **z** direction, with a corresponding drop of about 5 cm due to gravity at the recombination point. Ramsey fringe techniques can be used for detection by adapting the above scheme so that the final state is |c⟩ instead of |a⟩. Briefly, this involves omitting the **x** directed pulses that previously transferred the |c,0⟩ component to |a,0⟩. Instead, at the recombination point, a third set of 2N-1=49 pulses in the **z** direction will be used to create an equal superposition of |c,0⟩ and |b,+2ℏk⟩. Finally, to generate the Ramsey fringes, a pair of opposite circularly polarized pulses, counterpropagating along the **z** axis, will now be applied, in order to excite a Raman transition coupling the |c,0⟩ and |b,+ 2ℏk⟩ states (dotted-line transitions in Figure 1). As before, these pulses will be highly off-resonant with respect to the excited states of the $D_2$ manifold, with a two photon detuning denoted by Δ. For Δ=0, we choose a pulse duration correspond to a π/2 pulse coupling |c,0⟩ and |b,+2ℏk⟩. This π/2 pulse will transfer all the atoms from the superposition (|c,0⟩ + |b,+2ℏk⟩)/√2 into the |b,+102ℏk⟩ (and not the |c,0⟩ state, because of the sign difference introduced between the |a⟩ and |b⟩ states during each step of adibatic transfer). As Δ is scanned while the population in state |c⟩ is observed, a Ramsey type fringe would be seen, with a frequency width corresponding to 1/2πτ, where τ is approximately equal to the time separation between the first pulse that split the atoms into the two internal states, and the last pulse that recombines them. In this case, this linewidth will be about 2 Hz, corresponding to τ of about 102 msec. Of course, any systematic phase shift between the two arms will be manifested as the corresponding shift in the Ramsey fringe pattern, thus enabling detection of effects such as rotation.

Thus far, only one dimensional beamsplitters have been considered. However, for lithographic applications such as quantum dots, it is desirable to be able to split the atomic beam in two orthogonal directions, producing four components, which would yield a two-dimensional pattern. The scheme discussed above does not render itself easily to a two-dimensional generalization. Instead a modified scheme can be used. Again we assume that right after the atoms are released from the trap, they are each in state |a⟩, now denoted by |a, p=0,q=0⟩≡|a,0,0⟩, where p corresponds to momentum in the +**z** direction , and q corresponds to momentum in the +**x** direction. We now apply, along the **z** axis, two counter-propagating, σ+ polarized beams, exciting the transition between |a,0,0⟩ and |c,-2ℏk,0⟩. This is similar to the dashed-line π transitions in Figure 1, except that the excited state corresponds to $m_{F'}=0$. As before, these laser beams are assumed to be off-resonant with respect to the excited levels of the $D_2$ manifold, but two-photon resonant for this transition. Note that the beams have distinct frequencies, one (A) coupling |a⟩ to an excited state, while the other (C) is of a lower frequency, coupling |c⟩ to the same excited state. Unlike the case of adiabatic transfers, the pulses (each of the same duration) are now timed such that they both appear at the atom at the same time, thereby leaving at the same time. First, the time duration is chosen to correspond to a π/2 pulse area, with A propagating in the –**z** direction, while C propagating in the +**z** direction, so that the atom ends up in an equal superposition of |a,0,0⟩ and |c, -2ℏk,0⟩. This is illustrated in Figure 5 (solid line transitions). For concreteness, let us call this duration T'/2. The subsequent pulse pair is now assumed to be the same as before, except with a duration of T', and the directions of A and C reversed. This will cause a π pulse transition between |a⟩ and |c⟩, producing an equal superposition of |c,+2ℏk,0⟩ and

$|a,-4\hbar k,0\rangle$. This is illustrated in Figure 5 (dashed-line transitions). Note that the $\pi$ pulses excite two Raman transitions in parallel. Momentum selection rules ensure that there is no mixing of these transitions. The next pulse pair is identical to this one (i.e., of duration T'), except that the directions of A and C are again reversed. The state of the atom after this pulse will now be an equal superposition of $|a,+4\hbar k,0\rangle$ and $|c,-6\hbar k,0\rangle$. This is illustrated in Figure 5 (dotted-line transitions). After 2P number of such alternating pulse pairs (each pair with a duration T'), the atom will be in an equal superposition of $|a,+4P\hbar k,0\rangle$ and $|c,-(4P+2)\hbar k,0\rangle$.

Consider a case where 2P=24, so that after the sequence of pulses, the two components will differ in momentum in the **z** direction by 98 $\hbar$k, and separate out with a velocity of about 60 cm/sec. Thus, as before, the cloud will split up in two parts, with a separation of 2 mm after 3.3 msec. For this value of 2P, the two clouds correspond to states $|a, +48\hbar k,0\rangle$ and $|c, -50\hbar k,0\rangle$, respectively. We now apply a set of right circularly polarized pulses, with frequencies A and C, and propagating in the opposite directions, along the **z** axis. By choosing a pair of pulses, properly sequenced in directions, and numbering 2P=48, we can now reverse the direction of each of the components, producing an equal superposition of states $|a, -48\hbar k, 0\rangle$ and $|c, +46\hbar k, 0\rangle$.

While the spatially separated components of the superposition state are moving toward each other, we apply a pair of linearly polarized beams, co-propagating along the **x** direction, causing a Raman transition between $|a\rangle$ and $|c\rangle$ (dashed-line transitions in Figure 1)[16]. The duration and spatial location of this pulse pair is chosen such that a $\pi$ pulse is induced on the two-photon transition coupling $|a\rangle$ and $|c\rangle$, but only on the component of the cloud that corresponds to the state $|c,+46\hbar k,0\rangle$. The atom is now in an equal superposition of the states $|a,-48\hbar k,0\rangle$ and $|a,+46\hbar k,0\rangle$,[17] since the co-propagating fields give no net momentum transfer in the **x** direction.

While these two components are converging to each other in order to produce interference fringes in the **z** direction, we split (and later recombine) each component further along the **x** axis. Explicitly, we first apply a pair of linearly polarized beams, with frequencies A and C respectively, counter-propagating in the **x** direction. In a manner analogous to the **z** directed splitting, we first apply a $\pi/2$ pulse, interacting with both components of the split cloud, which produces an equal superposition of four states: { $|a, -48\hbar k,0\rangle$, $|c, -48\hbar k,-2\hbar k\rangle$} separated spatially in the **z** direction from { $|a, 46\hbar k,0\rangle$, $|c, 46\hbar k,-2\hbar k\rangle$}. This is followed by a series of direction-alternating $\pi$ pulse pairs, numbering 2Q, producing a set of four states:
{ $|a, -48\hbar k,4Q\hbar k\rangle$, $|c, -48\hbar k,-(4Q+2)\hbar k\rangle$} and { $|a, 46\hbar k,4Q\hbar k\rangle$, $|c, 46\hbar k,-(4Q+2)\hbar k\rangle$}.
The two subclouds in curly brackets are spatially separate from each other in the **z** direction, while inside each subcloud two sub-subclouds will now separate out in the **x** direction, with a velocity of 120 cm/sec if 2Q=4P=48. Thus, after about 1.7 msec, the separation in the **x** direction will be about 2 mm in each subcloud. At this point, a set of pulses totalling 2Q=96 will be applied in the **x** direction, with the timing sequences chosen so as to reverse the direction of splitting in the **x** direction. This will now produce an equal superposition of the following four states:

{ |a, -48ℏk,-96ℏk>, |c, -48ℏk,94ℏk>} and { |a, 46ℏk,-96ℏk>, |c, 46ℏk,94ℏk>}.

Finally, a **z** directed pair of copropagating, circularly polarized beams are now used to excite a π transition between |a> and |c>, but located spatially so as to affect only the |c> sub-subcloud inside each **z** separated subcloud. The spatial separation of 2 mm in the **x** direction makes this selective excitation possible. After this pulse sequence, we form four pieces of clouds, converging to one another in both **x** and **z** direction, and each in the internal state |a>:

{ |a, -48ℏk,-96ℏk>, |a, -48ℏk,94ℏk>} and { |a, 46ℏk,-96ℏk>, |a, 46ℏk,94ℏk>}.

Note that the clouds are now separated in the **z** direction by 1 mm, and in the **x** direction by 2 mm. Similarly, the speed of convergence in the **z** direction (about 60 cm/sec) is half that of the convergence speed in the **x** direction. As such, all four components of the cloud will come together in another 1.7 msec, forming a 2 dimensional matter wave grating pattern. Figure 6 shows schematically the splitting and recombining patterns. The bottom of figure 6 also shows the results of a numerical simulation, producing a two dimensional grating. The spacing of these patterns are determined by the values of P and Q: the peak to peak separation in the **z** direction is given approximately (for the rubidium transition wavelength of about 800 nm) by 100/P nm, and the separation in the **x** direction is 100/Q nm. For the numbers chosen here, we thus have a grating with about 4 nm spacing in the **x** direction, and 8 nm spacing in the **z** direction. Structures as small as 2 nm seem feasible given the source particles parameters considered here. The number of spots, and uniformity of height thereof, are determined largely by the coherence length of the sample. For a Bose condensed source, the coherence length is of the order of 300 μm, so that up to $10^{10}$ structures can be produced and deposited over an area of 300 μm diameter. In one dimension, these conclusions also apply to the adiabatic dark-state transfer interferometer discussed above.

A somewhat different approach can be used to produce two-dimensional structures with *arbitrary* patterns (as illustrated in figure 3). Briefly, The desired pattern (such as gears, turbines, cantilevers, etc.) is first drawn in a CAD program. The drawing is sampled as a two-dimensional function, f(x,y), from which one computes a new function: $g(x,y)=Cos^{-1}(f(x,y))$. A optical intensity mask is then produced, corresponding to g(x,y). Consider next the atomic wave. The atoms dropped from the magnetic trap (in the form of an atom laser) is first split, using a Raman resonant pulse, into two internal states. Both internal states are then defocused using a far-red-detuned laser beam with an anti-gaussian profile; this beam is pulsed on for a short time, then turned off. The expanding atomic waves are then collimated using another laser pulse with a gaussian profile. A third pulse, on resonance, carrying the planarized intensity pattern, then interacts with only one internal state of the atoms. For a short interaction time, the laser intensity pattern acts as a linear phase mask for the atomic wave. Both internal states are then defocused and recollimated. At this point, another Raman resonant pulse is used to convert all the atoms into the same internal state, so that they can interfere. The interference pattern is Cos(g(x,y)), which yields the original pattern, f(x,y). However, this pattern is now on a scale much shorter than the optical wavelengths. For parameters that are easily accessible, in the case of rubidium atoms, it should be possible to produce patterns with feature sizes of as small as 10 nm. As mentioned above, these patterns can be transferred to semi-conductors or coinage metals using chemical substitution techniques. Several layers can be bonded together to yield three dimensional structures, as is often done in current MEMS processes.

In summary, we have shown via explicit analysis as well as numerical simulation the design of two large angle interferometers. The first one uses the technique of adiabatic following in a dark state to produce a very large splitting-angle atomic interferometer, as well as one dimensional gratings as small as 2 nm spacing. This may lead to nearly two orders of magnitude improvement in the sensitivity of devices such as atomic gyroscopes, which are already as good as the best laser gyroscopes. The second one uses the technique of Raman pulses to produce a two-dimensional interferometer, with independent choice of grating spacings in each direction, each being as small as 2 nm. This scheme may enable us to produce uniform arrays of quantum dots with dimensions only 2 nm on each side. Finally, we have shown a novel scheme whereby it may be possible to generalize this process to produce arbitrary patterns with the same type of resolution, with applications to integrated circuits.

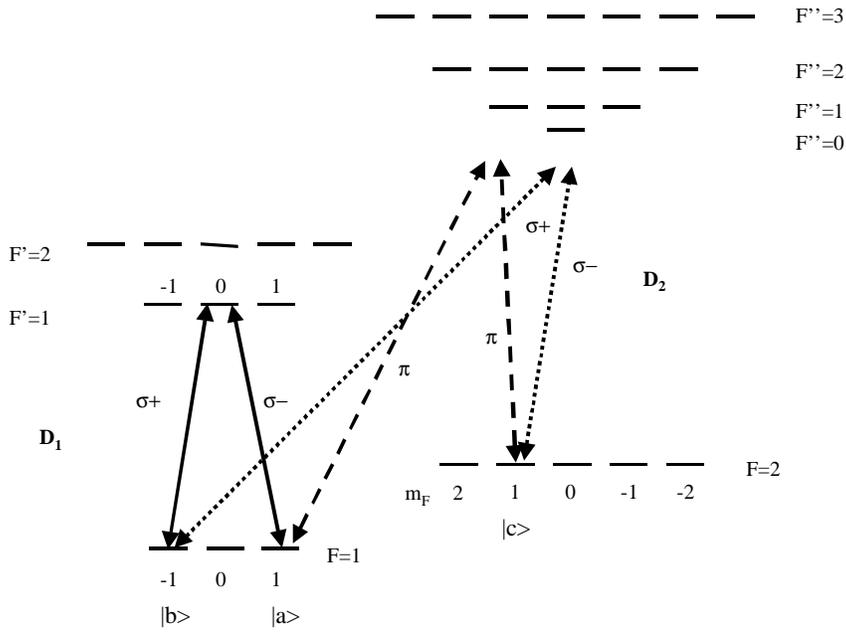

Figure 1. Schematic illustration of the relevant energy levels of $^{87}$Rb atoms, considered in this article as an example. Transitions from both the $D_1$ and the $D_2$ manifolds are used. The presence of two different types of Raman transitions in the $D_2$ manifold, excitable by optical beams propagating in orthogonal direction, is a key element of this design.

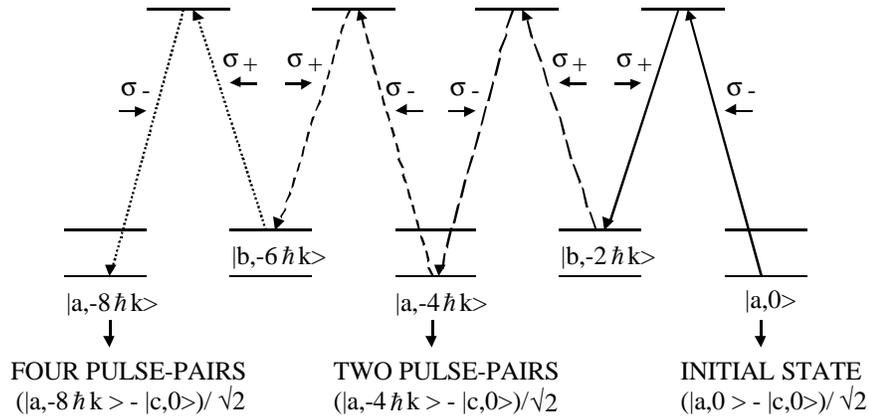

Figure 2. Schematic illustration of the first four pulse-pairs in the beamsplitter employing adiabatic transfer in the Raman dark state. Explicit form of the initial superposition state (after excitation with the **x** direction pulses) is shown along with the superposition states resulting after two and four **z** direction pulse-pairs are applied. Solid lines denote transitions excited by the first pair of pulses, long dashed lines denote the second pair, short dashed lines denote the third, and dotted lines the fourth. For clarity, the σ+ and σ- transitions are drawn as if they have different energies and energy shifts due to kinetic energy are omitted.

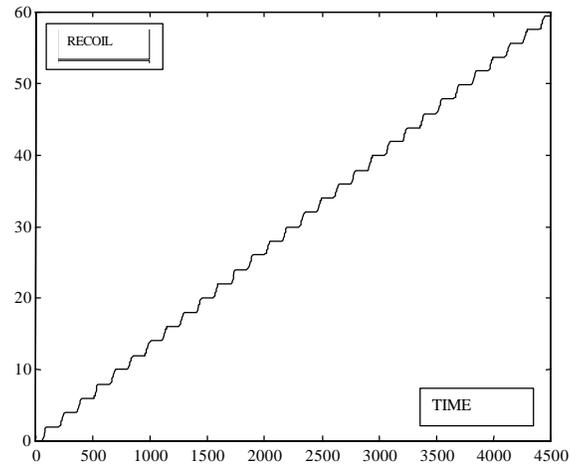

Figure 3. Result of numerical simulation, showing splitting corresponding to the abosorption of 60 recoil momenta. The process, compensated for Doppler detuning, is expected to continue upto a few hundred recoils until deleterious effects of imperfect adiabatic transfers would become noticeable.

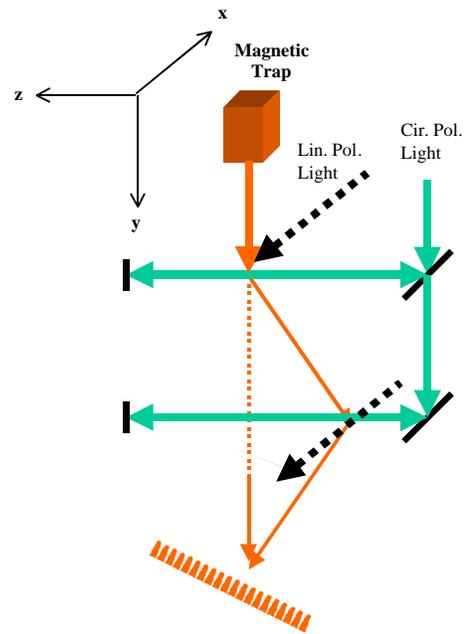

Figure 4. Schematic illustration of the three dimensional geometry employed in producing a beam splitter and recombiner via adiabatic following in the Raman dark state. For simplicity, this illustration assumes a constant downward velocity. The deflected component would follow a curved path in a gravitational field.

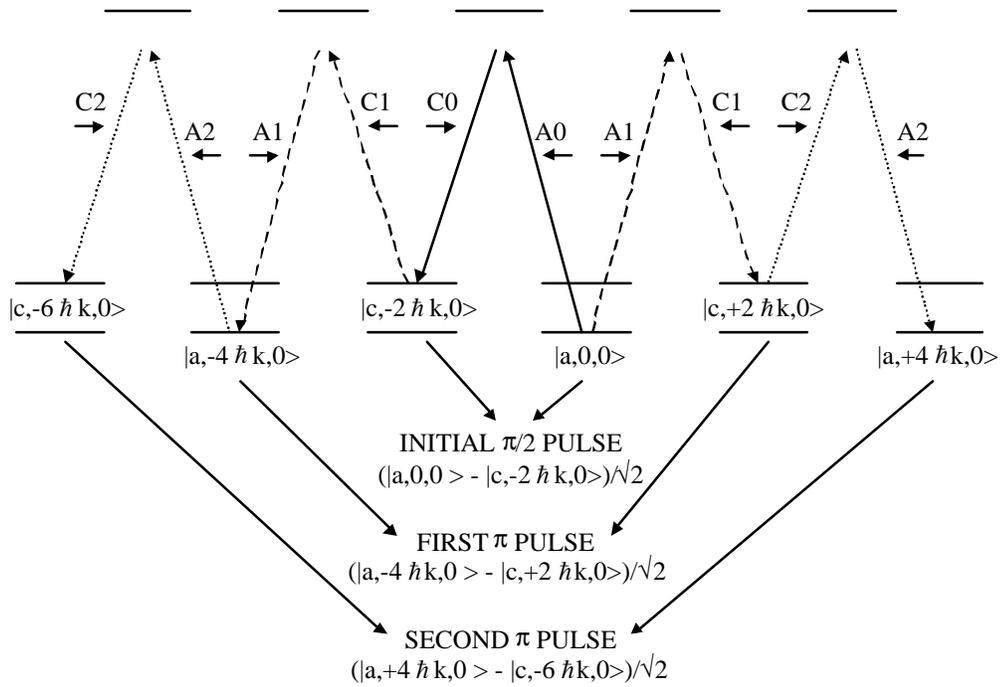

Figure 5. . Schematic illustration of the first three pulses in the Raman pulse beamsplitter. Explicit form of the initial superposition state, after excitation with the π/2 pulse, is shown along with the superposition states resulting after the first and second π pulses are applied. Solid lines denote transitions excited with the π/2 pulse, dashed lines denote the first π pulse, dotted lines denote the second π pulse. Note that the π pulses excite two Raman transitions in parallel. Momentum selection rules ensure that there is no mixing of these transitions. For clarity, the energy shifts due to kinetic energy are omitted.

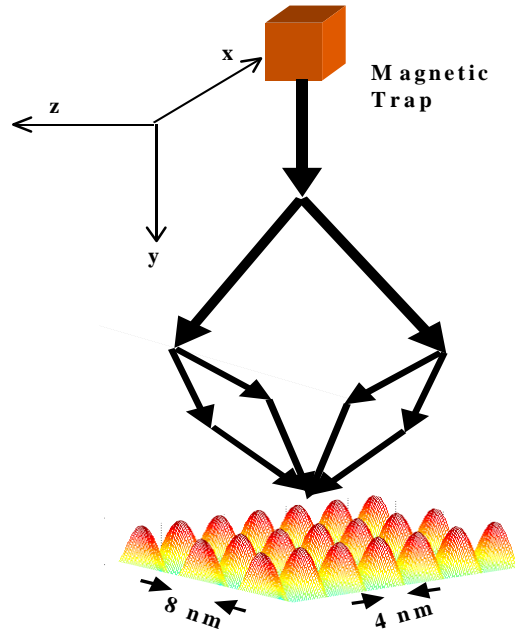

Figure 6. Basic illustration of the steps involved in producing two dimensional beam-splitting and recombining. For simplicity, the laser beams are not shown in the diagram. The two dimensional pattern at the bottom is produced via numerical simulation of the process described in the body of the text. For simplicity, this illustration assumes a constant downward velocity. The deflected components would follow a curved path in a gravitational field.

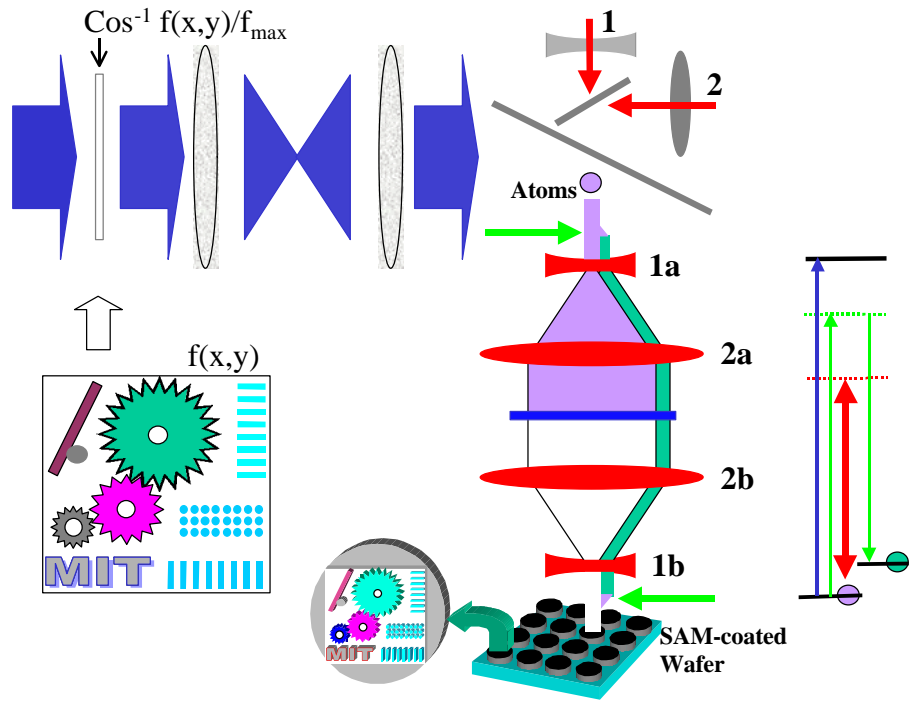

Figure 7. Basic illustration of the steps involved in producing two dimensional arbitrary patterns using a combination of atom focusing/defocusing and interferometry. Here, the inverse cosine of the desired pattern is first transferred to an optical intensity mask, which in turn acts as a phase mask (via ac-stark effect) for the atomic wave packet. See the body of the text for additional details.